\documentclass[12pt]{iopart}

\usepackage{array}
\newcolumntype{P}[1]{>{\centering\arraybackslash}p{#1}}

\usepackage{makecell}
\usepackage{float}
\usepackage{amssymb}
\usepackage{bm}
\usepackage{graphicx}
\usepackage{multirow}
\usepackage{dcolumn}

\expandafter\let\csname equation*\endcsname\relax
\expandafter\let\csname endequation*\endcsname\relax

\usepackage{amsmath}

\begin{document}

\newcommand{\mb}{{\bf m}}
\newcommand{\nb}{{\bf n}}

\title[Families of bosonic  suppression laws  beyond   the permutation  symmetry principle]{Families of bosonic  suppression laws  beyond   the permutation  symmetry principle}

\author{M. E. O. Bezerra and V. S. Shchesnovich}

\address{Centro de Ciências Naturais e Humanas, Universidade Federal do ABC, Santo André,  SP, 09210-170 Brazil }
\ead{matheus.eiji@ufabc.edu.br}
\vspace{10pt}
\begin{indented}
\item[]June 2023
\end{indented}

\begin{abstract}
Exact cancellation of quantum amplitudes in  multiphoton interferences with Fock states at input,  the so-called suppression or  zero transmission laws generalizing the Hong-Ou-Mandel dip, are useful tool  in quantum information and computation.   It was recently suggested that all  bosonic  suppression  laws follow from  a   common  permutation symmetry  in the input  quantum state   and   the unitary matrix   of   interferometer.   By using  the recurrence relations  for  interference of Fock states, we find a wealth of    suppression  laws on  the  beamsplitter and tritter  interferometers which do not follow from   the    permutation symmetry principle.    Our results reveal  the existence of whole    families  of   suppression laws for arbitrary  total  number of bosons  with only a fraction of them  being accounted for by  the  permutation symmetry principle,   suggested as  the general principle behind  the suppression laws. 
\end{abstract}

\section{Introduction}

 One of the most distinctive features of quantum theory is the superposition principle which, under appropriate conditions,  leads to the existence of  totally destructive interference in multi-path scenario,  with the probability of some   outcomes   being   exactly zero. When two   single photons become indistinguishable they bunch at the output of a balanced beamsplitter \cite{HOM}, which is the consequence of destructive interferences in the coincidence outcomes.      This is the well-known  Hong-Ou-Mandel   dip, which   has found numerous applications such as   characterization of photon indistinguishability \cite{Ou,IndPh},  generation and detection of  entanglement   \cite{Ent0,Ent1,Ent2} and    design of efficient quantum gates  \cite{OptGates} for all-optical computations. The exact cancellation can be understood as the consequence of a symmetry in the setup: the beamsplitter is balanced and the  Fock state of indistinguishable photons  is symmetric under the  transposition  of the input modes.    

The  totally destructive multiphoton interference for more than two  photons  has been studied    in many subsequent works,   including  the   even-odd number suppression events  and four-photon enhancement on a beamsplitter   \cite{BS_campos,Exp_4phBS},  the Hong-Ou-Mandel  type  effect  in the coincidence  counting   on the   symmetric Bell  (a.k.a. Fourier) multiports  \cite{HOM_Bell},  for which  the conditions for  all possible zero transmission laws were formulated    \cite{Tichy1} and  generalized   to both bosons and fermions \cite{Tichy2},   followed by a series of experiments with  various numbers of photons \cite{ExpFour1,ExpFour2,ULO,SYMDTh,3phPh,Gen3Ph}.  These works   pointed on a connection between  the  suppression  laws  and  some  underlying symmetry in the setup.   Such a connection was formulated   as  one common symmetry  principle  \cite{Dittel1,Dittel2}, which seemed to  explain  all  the known suppression laws, for bosons and fermions,  and generalize them to a wide class of   unitary interferometers (a.k.a. multiports) and input states.  

In present  work we reveal  the existence of families of   suppression  laws in   interference  with Fock states on  unitary multiports for arbitrary  total number of bosons, which are  not accounted for  by the  common permutation symmetry principle, suggested  previously as the  general principle behind  the suppression laws.

\section{Generating function and recurrence relations for  quantum  amplitudes}

Let $\hat{a}^\dagger_k$ and $\hat{b}^\dagger_l$ be respectively the creation operators  of optical mode in input port $k$ and optical output mode $l$  of a unitary   multiport of size $M$, with $k, l = 1, \ldots, M$. The output modes are related to the input modes by an  unitary multiport $U$ as follows
\begin{equation}
a^\dagger_k = \sum^M_{l=1} U_{kl} b^\dagger_l.
\label{expansion}
\end{equation}
We are interested  in the  $N$-photon  quantum amplitude between two Fock states ${}_b\langle  \nb  |\mb\rangle_a$, where $|\mb\rangle_a  = 
|m_1,\ldots,m_M\rangle= \prod_{k=1}^M \frac{(a^\dagger_k)^{m_k}}{\sqrt{m_k!}}|0\rangle$ and $|\nb\rangle_b  = 
|n_1,\ldots,n_M\rangle= \prod_{k=1}^M \frac{(b^\dagger_k)^{n_k}}{\sqrt{n_k!}}|0\rangle$, which is    proportional to   the matrix permanent  of a submatrix of $U$  \cite{origBS,BSAA},  i.e., a multilinear function of  the columns  and rows of the multiport matrix $U$ occupied by photons.  We will employ   the recurrence  relations satisfied by the matrix permanent, which follow from the generating function method (see for instance  Refs. \cite{Minc,Jackson}).   We  start  by observing that    $N$-photon  quantum amplitude    between  two Fock states   has  also a very  interesting   statistical interpretation   \cite{DG_FY}.   Assume that each photon ``possesses"   two independent properties  $(k,l)$ (a fictitious label):  the input port number it comes from, $k$,  and the output port number, $l$, where it lands.  Let  the entries of  $M\times M$-dimensional matrix $S$  give   a partition   of  $N$ photons by the two properties   ($S$ is called  contingency table in statistics).   The Fock state  amplitude  ${}_b\langle  \nb  |\mb\rangle_a$ is  proportional to the statistical average over  the  contingency tables $S$ with fixed margins, $m_k= \sum_{l=1}^M S_{kl}$ and $n_l= \sum_{k=1}^M S_{kl}$,     \cite{assymp1}:
\begin{equation}
{}_b\langle  \nb  |\mb\rangle_a = \frac{N!}{ \sqrt{\mb!\nb!}} \sum_{\{S \}} \mathcal{P}(S |\mb,\nb) \prod_{k=1}^M\prod_{l=1}^M U_{kl}^{S_{kl}},
  \label{FYA}
\end{equation}
where  $\mb!\equiv m_1!\ldots, m_M!$  and $ \mathcal{P}(S |\mb,\nb) $ is   the    Fisher-Yates distribution for  two independent properties~\footnote{Indeed,  the  multinomials $\binom{N}{ \mb}$, $\binom{N}{ \nb}$ and $\binom{N}{S }$  give, respectively,  the number of choices of $N$  photons for the input configuration, the output configuration, and for  a table    with given margins.} \cite{DG_FY},
\begin{equation}
\mathcal{P}(S |\mb,\nb) = \frac{\binom{N}{S }}{\binom{N}{ \mb}\binom{N}{ \nb}}=\frac{1}{N!}\prod_{k=1}^M\prod_{l=1}^M\frac{ m_k!n_l!}{S_{kl}!}.
\label{FY}
\end{equation}
 It is known that counting even the total number of  large-size tables with fixed margins     is a hard computational problem   \cite{DG_FY}, in agreement with the hardness of the quantum amplitude \cite{BSAA}. The  averaging   in Eq. (\ref{FYA}) over   the  tables with   fixed  margins    can  be cast in the form of partial derivatives of some  generating function.      Introducing the   formal variables, $x_1,\ldots, x_M$, we have   
      \begin{equation}
_b\langle { \bf n } | { \bf m } \rangle_a = \left. \prod^M_{l=1} \frac{1}{\sqrt{ n_l!}} \frac{\partial^{n_l}}{\partial x_l^{n_l} } G_{{\bf m}}( {\bf x}  ) \right|_{{\bf x} =0},
\label{rec}
\end{equation}
 with the generating function 
 \begin{equation}
G_{{\bf m}}( {\bf x} )  =  \prod^M_{k=1} \frac{1}{\sqrt{m_k!}} \left(  \sum^M_{l=1} U_{kl} x_l  \right)^{m_k} .
\label{Gen_fun}
\end{equation}
Indeed, the multinomial expansion of  each sum over $l$ in Eq. (\ref{Gen_fun})  introduces a table    $S$ satisfying $\sum_{l=1}^M S_{kl}= m_k$, whereas  taking the  derivatives   enforces the other margin, $\sum_{k=1}^M S_{kl} = n_l$, i.e.,  one   recovers the quantum amplitude in the form of Eq. (\ref{FYA}) (see also \ref{sup_func} for  more details).  

 The  expression in Eq. (\ref{Gen_fun}) admits some recurrence relations for the generating function with different total number of photons $N$. For instance,  taking one derivative over $x_l$ we get 
\begin{equation}
\frac{\partial}{\partial x_l} G_{{\bf m}} ({\bf x}) = \sum^M_{k=1} \sqrt{m_k} U_{kl} G_{{\bf m} - {\bf 1}_k} ({\bf x}) ,
\label{recurrence_amp}
\end{equation}
where $ {\bf m} - {\bf 1}_k \equiv (m_1,\ldots,m_k-1,\ldots,m_M)$ is the input configuration with one photon removed from the $k$-th mode.  The above generating function approach and the  expansion  in Eq. (\ref{FYA})  is intimately connected to  canonical transformations in the phase space  \cite{assymp2}.  The corresponding  recurrence relation for the amplitudes   can be obtained by replacing Eq.(\ref{recurrence_amp}) in  Eq. (\ref{rec}), which is the one derived in  Ref. \cite{Quesada}.  In addition,  another type of recurrence in  the two-mode case  for the quantum  probabilities, instead of the quantum amplitudes, was derived in Ref. \cite{Jabbour}.

\begin{figure}
\centering
\includegraphics[width=0.7 \textwidth]{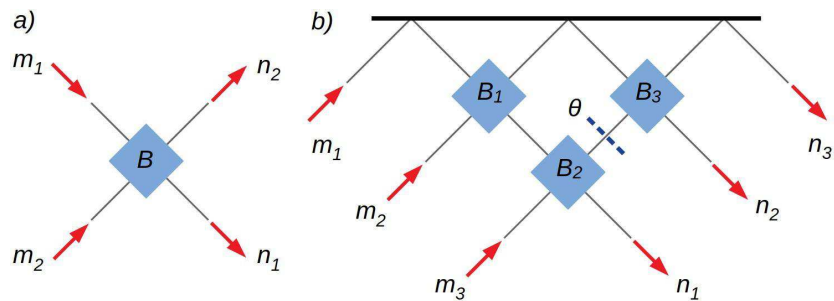}
\caption{Representation of the two interferometers that are considered to exemplify our method: a) Beamsplitter, that transforms two input modes into two output modes; b) Tritter, that is a composition of three different beamsplitters $B_1$, $B_2$, $B_3$ and a control phase shifter $\theta$. Here, each $m_k$ denotes the number of photons in the input mode $k$ and $n_l$ denotes the number of photons in the output mode $l$.}
\label{interferometers23}
\end{figure}

Let us now  focus on a single output port $l=1$, setting ${\bf n}=(n_1, {\bf n}_S)$, where  ${\bf n}_S=(n_2,...,n_M)$. Note that each derivative over $x_l$ in Eq. (\ref{rec}) removes a photon in the output $l$. Then, reusing the recurrence relation  of  Eq. (\ref{recurrence_amp})  repeatedly $n_l$ times for the output modes $l=2,\ldots,M$  we remove all the photons in this output mode, obtaining the amplitude $_b\langle {\bf n} | {\bf m} \rangle_a$   as a linear combination of the  amplitudes \mbox{$_b\langle n_1,{\bf 0}_S | {\bf m}' \rangle_a$}, where ${\bf m}' $ is the input configuration with fewer photons. The latter are simple enough  to be calculated directly. 
In the end we get    the amplitude in the form 
\begin{equation}
_b\langle {\bf n} | {\bf m}  \rangle_a = \sqrt{ \frac{n_1!}{{\bf n}_S! \mb! }} \left( \prod^M_{k=1} U^{m_k- |{\bf n}_S|}_{k1} \right) f^{\bf n}_{\bf m}(U),
\label{Form}
\end{equation}
where  $f^{\bf n}_{\bf m}(U) $ is a polynomial in the matrix elements of $U$, which we call as   the   suppression function and contains the  zero transmission laws  as being their  roots (see more details in \ref{sup_func} ).   Below we restrict ourselves to small numbers of photons in $M-1$ output  ports (i.e., the power of the polynomial $f^{\bf n}_{\bf m}(U) $  in Eq. (\ref{Form})), setting  $|{\bf n}_S|=1,2$ and illustrate our method on beamsplitter and tritter, given in      Fig. \ref{interferometers23}. 

We say that there is a  ``family of suppression laws" on the $M$-dimensional interferometer if  for  the  input $\mb$ and output $\nb$ configurations   of a given   form   and an  arbitrary compatible   total  number of bosons   there is a suppression law for  the  input and output configurations in such a form. 

\subsection{Families of suppression laws on the    beamsplitter}

 Let us first test the method using the  beamsplitter, illustrated in Fig.\ref{interferometers23}(a), with the matrix
\begin{equation}
B   =
\begin{pmatrix}
\sqrt{\tau} & -\sqrt{\rho} e^{-i \varphi} \\
\sqrt{\rho} e^{i \varphi} & \sqrt{\tau}
\end{pmatrix}
\label{beamsplitter}
\end{equation}
where $\tau$ is the transmissivity, $\rho=1-\tau$ the reflectivity and $\varphi$ is the reflection phase.   For now, we can  neglect this reflection phase as it can be scaled out, however, when considering the tritter decomposition, as in Fig. \ref{interferometers23}(b), this phase is an important parameter.   In this case  ${\bf n}_S = n_2$. The   beamsplitter of Eq.(\ref{beamsplitter}) with arbitrary $\tau$  is also the  composition of two balanced beamsplitters and two additional phase shifters between them, in such a way that the  transmission parameter    $\tau$ is controlled by the phase shifters \cite{matrix1}.

For    $n_2=1$   the  recurrence  in Eq. (\ref{Form})   has the following function 
\begin{equation}
f^{(n_1,1)} _{(m_1,m_2)} (B)  =   (m_1+m_2) \tau - m_1,
\label{fun_b1}
\end{equation}
 implying   that   the  quantum amplitude   \mbox{${}_b\langle n_1, 1 | m_1 , m_2 \rangle_a=0$} for an  arbitrary $n_1\ge 1$  and the  transmission
\begin{equation}
\tau^{(1)}  = \frac{m_1}{m_1+m_2}.
\label{sup_BS_Jabbour}
\end{equation}
This   coincides with the  previous result    \cite{Jabbour}, obtained by another method. The whole  family of such suppression laws  contains also  the  HOM effect \cite{HOM} for the    symmetric beamsplitter  for $m_1=m_2=1$.  

For   $n_2=2$ we get the 
 suppression function 
\begin{eqnarray}
&& f^{(n_1,2)} _{(m_1,m_2)} (B) =    (m_1 + m_2 -1) (m_1+m_2) \times  \nonumber\\
&& \times \Bigl[  \tau^2  - \frac{2m_1}{m_1+m_2} \tau  + \frac{m_1(m_1-1)}{(m_1+m_2)(m_1 + m_2 -1)} \Bigr],
\label{fun_b2}
\end{eqnarray}
giving  another  (previously unknown)  suppression  law   $\langle n_1, 2 | m_1 , m_2 \rangle=0$ for the  transmission 
\begin{equation}
\tau^{(2)} =\frac{m_1}{m_1+m_2} \left(1  \pm \sqrt{\frac{ m_2/m_1 }{ m_1+m_2-1}}\right).
\label{sup_BS_two}
\end{equation}
This family of  suppression  laws   also  contains  the symmetric beamsplitter $\tau^{(2)}=1/2$ for specific inputs, e.g.,  for four input photons  ${}_b\langle 2,2|1,3\rangle_a=0$  \cite{BS_campos,Exp_4phBS}. Only such cases  can be explained by the permutation symmetry approach \cite{Tichy1,Tichy2,Dittel1,Dittel2} (in the above case the  transposition symmetry of  two output ports with $n_1=n_2=2$). 

\begin{figure}
 \centering
\includegraphics[width=0.5\textwidth]{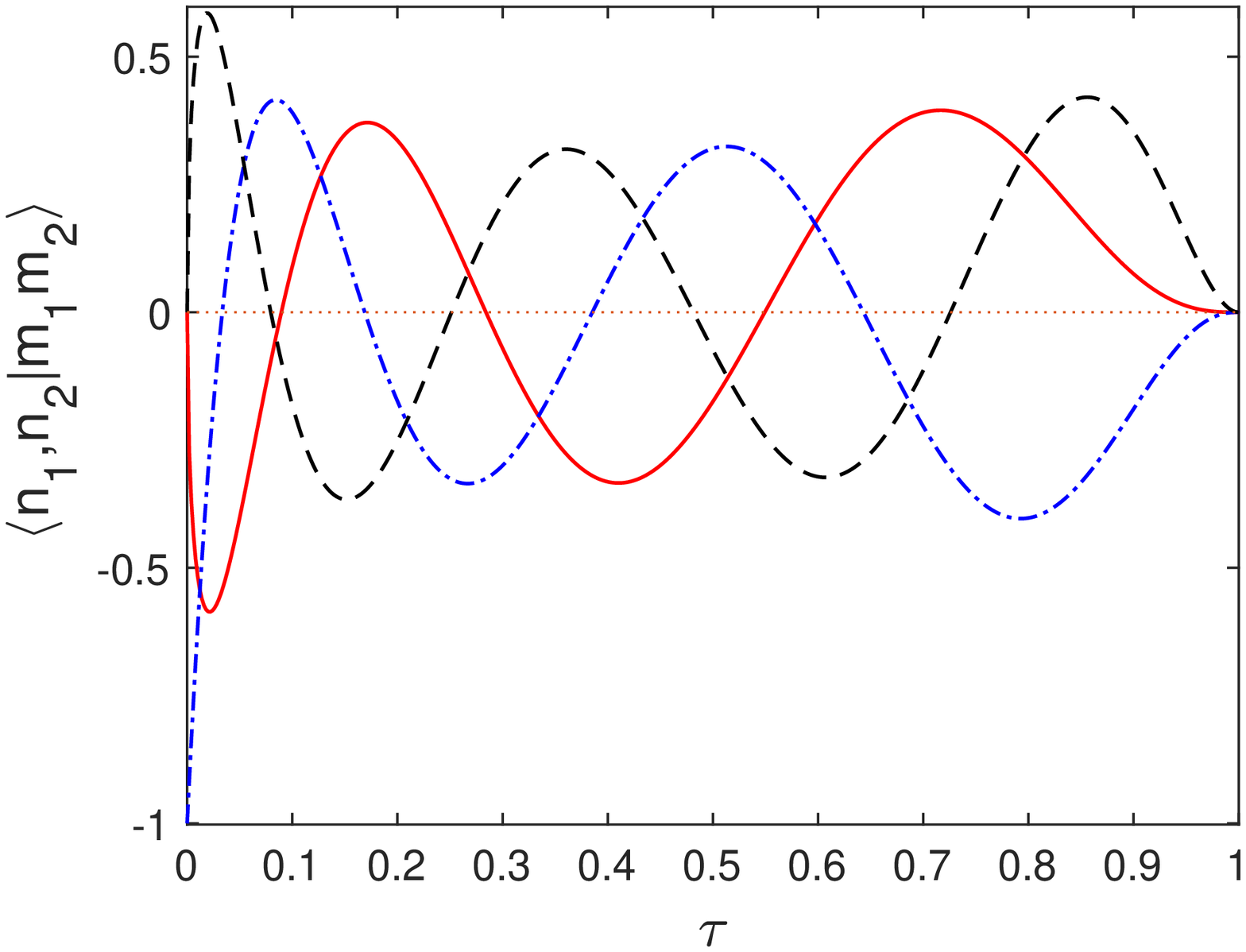} \caption{ Typical behaviour of  the quantum amplitudes on a beamsplitter  and the   interlaced zeros (the suppression laws). Here  we plot   ${}_b\langle n_1,n_2|9,4\rangle_a$ as functions of the beamsplitter transmission  $\tau$ for $n_1 = 3$ (solid line), $n_1 = 4$ (dash-dotted line), and $n_1=5$ (dashed line). }
 \label{FIG2}
 \end{figure}

 The above presented  approach  allows one   to derive  all possible  suppression laws for the beamsplitter.  The computations, however, become quite involved  as the  minimum  number of bosons in  the input and output ports scales up.  Nevertheless, some general conclusions  are allowed by the  fact that  the quantum amplitudes ${}_b\langle n_1,n_2 |  m_1,m_2 \rangle_a$ on  a beamsplitter  can be made  real-valued  functions of its  transmission $\tau$ by removing the overall phase.  Numerical simulations    with various distributions of bosons (i.e., Fock states)  reveal that  the  number of   zeros in   a quantum amplitude   is given by  the minimum number of bosons $\mathrm{min}(n_l,m_k)$  in the four ports.   Moreover, two quantum amplitudes   related by the  exchange of a single boson  have interlaced zeros: between two zeros of one of them there is one  zero of the other, see also Fig. \ref{FIG2}  (at the end points,  $\tau =0 $ and $\tau = 1$,   a real-valued quantum amplitude  can be  either equal to zero or to  $\pm 1$, which  explains the above  bound on the total number of zeros).  
 
\subsection{Families of suppression laws on the  tritter}  

We now consider the suppression laws  on the tritter   obtained by an arrangement of three beamsplitters   according to the setup in  Fig. \ref{interferometers23}(b)     \cite{tritter_campos, matrix1}.  Here,  each beamsplitter has a matrix  $B_j$  similar to that of Eq. (\ref{beamsplitter})  with   the  transmissivity $\tau_j $ and   phase $\varphi_j $. An additional phase plate   $\theta$ is inserted in one of the optical paths.    Our   tritter has  in total seven  free parameters,  hard to analyse  in the  general case.   We will therefore focus on  two specific  families each having   only two free parameters. For the first family  we set: $\tau_2 = 2/3$,  $\tau_3 =   1/2$, $\varphi_j =  \pi/2$, leaving us with the free parameters $\tau_1$ and $\theta$. It has the following matrix   
\begin{eqnarray}
&& T^{(1)}  =
\frac{1}{\sqrt{6}}
\begin{pmatrix}
2 \sqrt{\tau_1}, & - \sqrt{\tau_1} e ^{i \theta} - i \sqrt{3 \rho_1} ,& - \sqrt{\tau_1} e ^{i \theta} + i \sqrt{3 \rho_1} \\
2 \sqrt{\rho_1}, &  - \sqrt{\rho_1} e ^{i \theta} + i \sqrt{3 \tau_1} ,  &  - \sqrt{\rho_1} e ^{i \theta} - i \sqrt{3 \tau_1} \\
\sqrt{ 2}, &  \sqrt{2} e^{i \theta}, &  \sqrt{2} e^{i \theta}
\end{pmatrix}. \nonumber\\
\label{rho1_theta}
\end{eqnarray}
For the second family  we set: $\tau_1 =  \tau_3 = 1/2$ and $\varphi_j =  \pi/2$, with the free parameters being $\tau_2$ and $\theta$.  It has  the following matrix
\begin{equation}
T^{(2)}   = \frac{1}{2}
\begin{pmatrix}
\sqrt{2 \tau_2}, & -i - \sqrt{\rho_2} e^{i \theta}, & i - \sqrt{\rho_2} e^{i \theta} \\
\sqrt{2 \tau_2}, & i - \sqrt{\rho_2} e^{i \theta} ,& -i - \sqrt{\rho_2} e^{i \theta} \\
2 \sqrt{ \rho_2},& \sqrt{2 \tau_2} e^{i \theta}  ,& \sqrt{2 \tau_2} e^{i \theta}
\end{pmatrix}.
\label{rho2_theta}
\end{equation}
The above two tritter families reduce to  the well-known symmetric tritter (i.e.,  Bell multiport) when    $\theta=0$ and, in the first case, $\tau_1 = 1/2$  or, in the second case, $\tau_2 = 2/3$. 

\begin{figure}
\centering
\includegraphics[width=0.8 \textwidth]{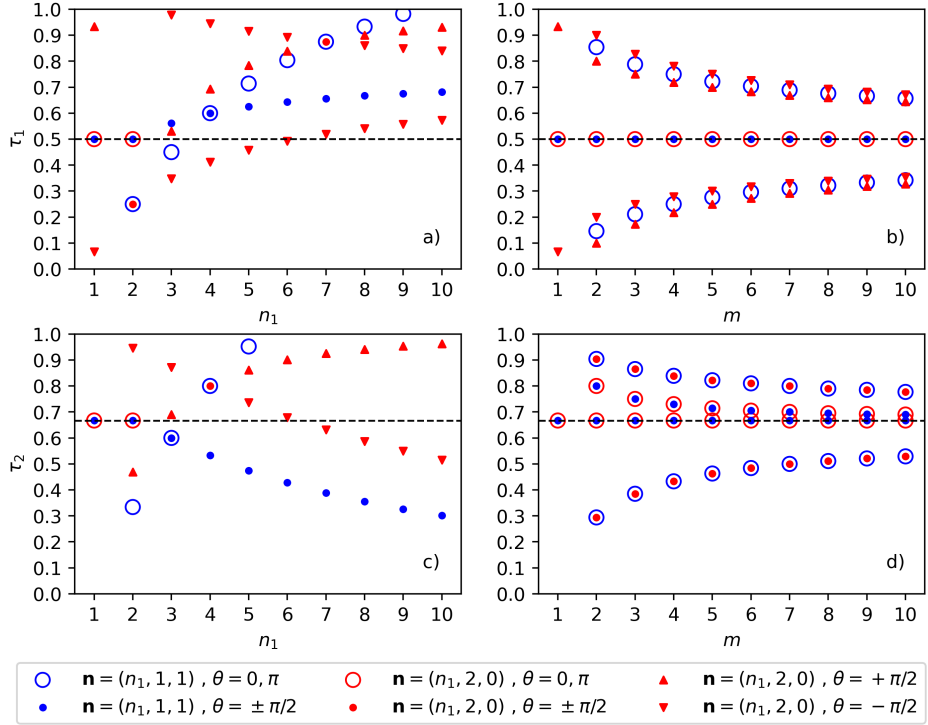}
\caption{Non-trivial suppression laws for outputs ${\bf n}=(n_1,1,1)$ and ${\bf n}=(n_1,2,0)$.  (The suppression laws for $\tau_j=0$ or $\tau_j=1$ are  trivial and were  removed from the graph.) For the tritter $T^{(1)}$   the suppression laws are  for the inputs: a) ${\bf m}^{(I)}=(n_1,1,1)$ and b) ${\bf m}^{(II)}=(m,m,m)$. For the tritter $T^{(2)}$   the suppression laws are for the inputs: c) ${\bf m}^{(I)}=(n_1,1,1)$ and d) ${\bf m}^{(II)}=(m,m,m)$. The dashed line corresponds to the symmetric tritter $\tau_1=1/2$ and $\tau_2=2/3$ for $\theta=0$.}
\label{Tritter_fig}
\end{figure}

For the  tritter, in contrast to the beamsplitter,   two   input mode occupations can vary  for  a given total number of bosons.  We will focus below  on  the following two particular families of input states  $\mb^{(I)}=(n_1,1,1)$ and $\mb^{(II)}=(m,m,m)$ with some $n_1\ge 1$ and $m\ge 1$.  This choice of specific inputs is also dictated by the need to compare with the suppression laws due to the  permutation symmetry principle. For $|{\bf n}_S|=2$ we have found  suppression  laws for the outputs $\nb=(n_1,1,1)$ and  $\nb=(n_1,2,0)$, which are shown in Fig. \ref{Tritter_fig}.  The explicit expressions for the corresponding  suppression functions  $f^{\bf n}_{\bf m}(T) $ and some of the suppression laws are presented in \ref{sup_func}.

\subsection{Suppression laws from the  permutation symmetry}

Only a fraction  of the suppression laws discussed above are explained by the ``general permutation symmetry   principle"  of Refs. \cite{Dittel1,Dittel2}, given by the red circles on the dashed line in Fig. \ref{Tritter_fig}), corresponding to the   input   $\mb^{(II)}$ and output $\nb=(n_1,2,0)$ (with $n_1=3m-2$). These appear for  the symmetric tritter, with the three-dimensional Fourier matrix  
\begin{equation}
T_s=  \frac{1}{\sqrt{3}}
\begin{pmatrix}
1 & -\frac{1+i\sqrt{3}}{2} &\frac{-1+i\sqrt{3}}{2} \\
1 & \frac{-1+i\sqrt{3}}{2} &- \frac{1+i\sqrt{3}}{2} \\
1 & 1 & 1
\end{pmatrix},
\label{Trs1}
\end{equation}
obtained  by setting either $\tau_1=1/2$ in  Eq. (\ref{rho1_theta}) or $\tau_2=2/3$ in Eq. (\ref{rho2_theta}) and $\theta=0$. Such  suppression laws     also are related to some symmetry of the  suppression function  in Eq. (\ref{Form}), in which the corresponding roots do not depend   on $n_1$  and  $m$. We can verify it from the following expressions:
\begin{equation}
f^{(n_1,2,0)}_{(m,m,m)} (T^{(1)})  \overset{\theta=0}{=} \frac{m}{27} \left[ 3(m-1)(2\tau_1-1)+2i \sqrt{3 (1-\tau_1) \tau_1} \right] (2 \tau_1-1) ,
\label{123_1_n20}
\end{equation}
\begin{equation}
f^{(n_1,2,0)}_{(m,m,m)}  (T^{(2)})  \overset{\theta=0}{=} - \frac{m}{8} \left[ (3m-1)\tau_2-2m  \right] (3 \tau_2 - 2) \tau_2 .
\label{123_2_n20}
\end{equation}
whose the constant roots $\tau_1=1/2$ and $\tau_2=2/3$ are related to the permutation symmetry principle and the other roots are the suppression laws outside the dashed line in Fig. \ref{Tritter_fig}(d), which cannot be explained by this principle.

Interestingly,   we have found a tritter  $\widetilde{T}_s$ satisfying a different type of symmetry.  This new tritter corresponds to a real (orthogonal)   matrix in a form similar to that of  $T_s$  in Eq. (\ref{Trs1}):
 \begin{equation}
\widetilde{T}_s =  \frac{1}{\sqrt{3}}
\begin{pmatrix}
1 &- \frac{1+\sqrt{3}}{2} & \frac{-1+\sqrt{3}}{2} \\
1 & \frac{-1+\sqrt{3}}{2} & -\frac{1+\sqrt{3}}{2} \\
1 & 1 & 1
\end{pmatrix} .
\label{Trs2}
\end{equation}
and is obtained by setting either $\tau_1=1/2$ in  Eq. (\ref{rho1_theta}) or $\tau_2=2/3$ in Eq. (\ref{rho2_theta}) and    $\theta=\pi/2$ (factoring out the unimportant total phases in the output modes).   It shares one of the symmetries with that of Eq. (\ref{Trs1}):   it is   invariant under the simultaneous  permutation of    rows $1$ and $2$ and  columns $2$ and $3$ (not the  same  symmetry as required by  the  ``general permutation symmetry   principle"  of  Refs. \cite{Dittel1,Dittel2} for the considered quantum amplitudes). In addition, this symmetric tritter in  Eq. (\ref{Trs2})    results from the   transposition operation   of the first and the third  inputs  ($P_{13}$), followed by  a balanced beamsplitter on the second and third inputs ($B(\tau_s)$),  and  then by the  inverse of the symmetric tritter $T_s$, i.e.,  we have  \mbox{ $\widetilde{T}_s =P_{13}\left(1\bigoplus B(\tau_s)\right)T^\dag_s$}, where the beamsplitter is given by Eq. (\ref{beamsplitter})  with $\tau_s = (\sqrt{3}+i)/4$.  

The  suppression laws  on  the symmetric  tritter $\widetilde{T}_s$ of Eq. (\ref{Trs2})  corresponding  to the    input   $\mb^{(II)}$ and output $\nb=(n_1,1,1)$ are given by the blue points on the dashed lines in Fig. \ref{Tritter_fig}.  However, these suppression laws    cannot be    explained by the  ``general permutation symmetry   principle"  of Refs. \cite{Dittel1,Dittel2} which is applicable only to the standard symmetric tritter $T_s$ .

In addition, denoting these symmetric tritters as $T_s (0) = T_s$ and $T_s (\pi/2) = \widetilde{T}_s$, is is easy to prove that Eqs.(\ref{Trs1}),(\ref{Trs2}) are obtained explicitly from the following matricial construction:
\begin{equation}
T_s (\theta) =
\begin{pmatrix}
-\frac{1}{\sqrt{2}} &  \frac{i}{\sqrt{2}} & 0 \\
\frac{i}{\sqrt{2}} &  - \frac{1}{\sqrt{2}} & 0 \\
0 & 0 & 1 
\end{pmatrix}
\begin{pmatrix}
\sqrt{\frac{2}{3}} & 0 & \frac{i}{\sqrt{3}} \\
0 & 1 & 0 \\
\frac{i}{\sqrt{3}} & 0 & \sqrt{\frac{2}{3}}
\end{pmatrix}
\begin{pmatrix}
1 & 0 & 0 \\
0 & 1 & 0 \\
0 & 0 & \text{e}^{i\theta}
\end{pmatrix}
\begin{pmatrix}
1 & 0 & 0 \\
0 & \frac{1}{\sqrt{2}} & -\frac{i}{\sqrt{2}} \\
0 & \frac{i}{\sqrt{2}} & -\frac{1}{\sqrt{2}} 
\end{pmatrix} ,
\end{equation}

We  have  also analysed the suppression function for  the  amplitudes $_b\langle n_1,1,0 | m,m,m \rangle_a$ and $_b\langle n_1,0,1 | m,m,m \rangle_a$. These amplitudes are zero only for the  symmetric tritters  $T_s$ and $\widetilde{T}_s$, as shown in \ref{apptritter}.  From   the permutation symmetry of Refs. \cite{Dittel1,Dittel2}  these suppression laws   follows  only for  the tritter $T_s$. 
    
\section{Suppression laws and  partial distinguishability}

 Photons are  partially distinguishable due to degrees of freedom not acted upon by the interferometer, which are called the  internal states.  In Ref.~\cite{DTh} it   has been conjectured that  the zero probability  in the  output of  multi-photon interference with partially distinguishable photons is invariably the result of an exact cancellation of the quantum amplitudes of only the completely indistinguishable photons. This conjecture generalizes the well-known HOM effect \cite{HOM} to more than two photons and arbitrary interferometer (also to non-ideal detectors) and  the observations made in Ref. \cite{TichyDth}. It has been  confirmed by all    suppression laws in    Refs. \cite{Dittel1,Dittel2}.  Thus, by the conjecture, any   suppression law which is not broken by partial the distinguishability of photons  needs  other    suppression laws for  smaller total numbers of photons.

Now, this effect will be illustrated by a simple case. Let an experimental setup where $N$ photons are prepared from independent sources  in   either $N$ pure internal states  $|\phi_i\rangle$, $i=1,\ldots, N$. If, for instance, an input has one mode occupied by one photon and this photon is partially distinguishable from the  rest of $N-1$ photons,   we can use   just two internal states $|1\rangle$ and $|2\rangle$, with $|\phi_{k}\rangle = |1\rangle$ for $1\le k\le N-1$ and $| \phi_N \rangle =\text{cos} \hspace{0.5mm} \alpha \hspace{0.5mm} | 1 \rangle + \text{sin} \hspace{0.5mm} \alpha \hspace{0.5mm} | 2 \rangle$.  Note that, the last photon becomes indistinguishable from the others when $\alpha=0$ and distinguishable when $\alpha=\pi/2$. Therefore, we have the following state at the input:
\begin{equation}
\hat{\rho}_{\bf m} = \frac{1}{\bf m!} \prod^{N-1}_{i=1} \hat{a}^\dagger_{k_i , 1}  \hat{a}^\dagger_{k_N , \phi_N}    | 0 \rangle \langle 0 |   \prod^{N-1}_{i=1} \hat{a}_{k_i , 1}   \hat{a}_{k_N , \phi_N}
,
\end{equation}
where the first index of the creation/annihilation operators is related to the spatial mode and the second index to the internal state. The creation operator of the $N$-th photon is then given by:
\begin{equation}
\hat{a}^\dagger_{k_N, \phi_N}=\text{cos}\alpha \hspace{0.5mm} \hat{a}^\dagger_{k_N, 1} + \text{sin}\alpha \hspace{0.5mm} \hat{a}^\dagger_{k_N, 2} ,
\end{equation}

We define a set of POVMs $\hat{\Pi}_{\bf n} $ related to the detection of the photons in the configurations ${\bf n}$ at the output:
\begin{equation}
\hat{\Pi}_{\bf n} =  \frac{1}{\bf n!} \sum_{\bm{j}} \prod^N_{i=1} \hat{b}^\dagger_{l_i , j_i} | 0 \rangle \langle 0 |  \prod^N_{i=1} \hat{b}_{l_i , j_i}
,
\end{equation}
where the sum in ${\bm j}$ is over the internal states $j_i=1,2$. Then, after some calculations and defining $ {\bf n} - {\bf 1}_l \equiv (n_1,\ldots,m_l-1,\ldots,n_M)$ , we can obtain the following expression for the probability:
\begin{eqnarray}
&& P({\bf n} | {\bf m},\alpha) = \nonumber\\
&& = \sum_{\bm{j}} \text{Tr}\left( \hat{\rho}_{\bf m} \hat{\Pi}_{\bf n} \right) \nonumber\\
&& =  \frac{1}{{\bf m}!{\bf n}!}  \sum_{\bm{j}} \left|
\langle 0 | \prod^N_{i=1} \hat{b}_{l_i , j_i} \prod^{N-1}_{i=1} \hat{a}^\dagger_{k_i , 1} \left( \text{cos}\alpha \hspace{0.5mm} \hat{a}^\dagger_{k_N, 1} + \text{sin}\alpha \hspace{0.5mm} \hat{a}^\dagger_{k_N, 2} \right) | 0 \rangle
\right|^2 \nonumber\\
&& = \frac{\text{cos}^2 \alpha}{{\bf m}!{\bf n}!}  \Big|
\langle 0 \left| \prod^N_{i=1} \hat{b}_{l_i , 1} \prod^{N}_{i=1} \hat{a}^\dagger_{k_i , 1} | 0 \rangle
\right|^2 +  \frac{\text{sin}^2 \alpha}{{\bf m}!{\bf n}!} \sum_{\bm{j}}  \left|
\langle 0 | \prod^N_{i=1} \hat{b}_{l_i , j_i} \prod^{N-1}_{i=1} \hat{a}^\dagger_{k_i , 1} \hat{a}_{k_N , 2} | 0 \rangle
\right|^2  \nonumber\\
&& = \text{cos}^2 \alpha \left| _b\langle {\bf n} | {\bf m} \rangle_a \right|^2 +  \text{sin}^2 \alpha \sum^M_{l=1} |U_{kl}|^2 \left| _b\langle {\bf n}- {\bm 1}_l | {\bf m}- {\bm 1}_k  \rangle_a \right|^2 .
\end{eqnarray}

In the previous equation, we have developed suppression laws for the amplitudes $_b \langle {\bf n} | {\bf m} \rangle _a$ in the main text. However, in principle, the other terms $_b\langle {\bf n}- {\bm 1}_l | {\bf m}- {\bm 1}_k  \rangle_a$ are non zero and then we need to use another sequence of recurrence relations to eliminate the photons at ${\bf n}- {\bm 1}_l$. Let us focus on the distinguishable projection of the previous equation. The sum over $l$  has  $M$ non-zero  terms, each one being a product of two probabilities:   a probability  of the transition of one  distinguishable photon to one   output mode $l$   (such that $n_l>0$ in $\nb$) multiplied by the probability of  detecting  the remaining $N-1$ indistinguishable photons  to  the reduced output $ \nb-\bm{1}_l$.   Except the trivial case  of the single-photon probability being zero, all probabilities of detecting $N-1$ photons in the outputs  $ \nb-\bm{1}_l$  should be zero for zero output probability of such $N$  photons. In the sequence, we will illustrate this effect for the interferometers considered.

\subsection{ Beamsplitter}

To illustrate this effect in our results, let us consider the simple example, where have $m_1+m_2$ photons interfering in a beamsplitter and we want to calculate the probability $P(n_1,1|m_1,m_2,\alpha)$. Considering that the partially distinguishable photon is injected at the input mode $k=1$, we arrive at the following probability:
\begin{eqnarray}
&& P(n_1,1|m_1,m_2,\alpha) = \nonumber\\ 
&& =  \text{cos}^2 \alpha | _b\langle  {\bf n} | {\bf m} \rangle_a |^2  +  \text{sin}^2 \alpha \Big( |U_{11}|^2 | _b\langle n_1-1,1|m_1-1,m_2 \rangle_a |^2 + \nonumber\\ && + |U_{12}|^2 | _b\langle n_1,0|m_1-1,m_2 \rangle_a |^2 \Big) ,
\end{eqnarray}
where the first term is zero for $\tau = m_1/(m_1+m_2)$, according to Eq.(\ref{sup_BS_Jabbour}). However, ignoring the trivial solutions $\tau=0,1$, the second term is zero when $\tau=(m_1-1)/(m_1+m_2-1) $ and the last is zero only for trivial solutions. Therefore the suppression law is broken,  as the probability $P(n_1,1|m_1,m_2,\alpha)$ is no longer zero, because the three terms cannot be simultaneously zero for $\tau \neq 0,1$.

\subsection{ Tritter}

Now, let us consider the interference in the tritter $T^{(1)}$, with phase $\theta=\pi/2$, and the probability $P(n_1,1,1|n_1,1,1,\alpha)$. If the partially distinguishable photon is injected at $k=1$, we have
\begin{eqnarray}
&&  P(n_1,1,1|n_1,1,1,\alpha) = \nonumber\\
&& =  \text{cos}^2 \alpha | _b\langle  {\bf n} | {\bf m} \rangle_a |^2 + \hspace{1mm} \text{sin}^2 \alpha \Big( |U_{11}|^2 | _b\langle n_1-1,1,1|n_1,1,0 \rangle_a |^2 + \nonumber\\
&& + |U_{12}|^2  | _b\langle n_1,0,1|n_1,1,0 \rangle_a |^2 + |U_{13}|^2  | _b\langle n_1,1,0|n_1,1,0 \rangle_a |^2 \Big) ,
\label{tritter_partial}
\end{eqnarray}
where the first term is zero for $\tau_1=3n_1/(4n_1+1)$, according to Table \ref{table1} in \ref{apptritter}. The other three need to satisfy respectively the following equations
\begin{eqnarray}
&& (n_1+1) \sqrt{\tau_1 (1-\tau_1)} + \sqrt{3}(n_1+1) \tau_1 - \sqrt{3} n_1 = 0,
\nonumber\\
&& (n_1+1) \sqrt{\tau_1 (1-\tau_1)} - \sqrt{3}(n_1+1) \tau_1 + \sqrt{3} n_1 = 0,
\nonumber\\
&& 4(n_1+1) \tau_1 \sqrt{1-\tau_1} - 3 \sqrt{1-\tau_1} = 0.
\end{eqnarray}
where the last lead to $\tau_1=1$ or $\tau_1=3/4(n_1+1)$, which are not solutions of the first two equations. Therefore, the probability $P(n_1,1,1|n_1,1,1,\alpha)$ cannot be zero.

\section{Conclusion}

 We have revealed the existence of   whole  families of  the suppression laws on the  beamsplitter and   tritter  multiports   for arbitrary total number of photons, which are not explained by the permutation symmetry principle advanced in Refs. \cite{Tichy1,Tichy2,Dittel1,Dittel2}.  We  have  discussed above  only  a fraction of all  possible  suppression laws  on the tritter,    numerical simulations reveal  additional   families  of the   suppression laws not related to the permutation symmetry  principle.   Similar    suppression laws, not explained by the permutation symmetry principle, are  expected to appear  for  multiports of any size and any total number of bosons,  since by using our generation function  approach one can, in principle,  obtain all   the  suppression laws for a multiport of any size (though this is impractical by the complexity of the calculations which    involve finding  roots of higher-order polynomials).    One can, on the other hand,  explore  the suppression laws    experimentally,  due to the  recent   breakthrough in the  controlled production of     Fock states with  specified  number of photons: by using heralded Fock states from a SPDC process \cite{Fock1},  the interaction of  a coherent state with two-level atoms \cite{Fock2}, and by converting a coherent state into a Fock state inside a resonator  by    radiation losses \cite{Fock3}. Our results also  beg the important general question:  Can the   discovered   families of  suppression laws      follow from   a  yet more general common  symmetry principle?  This could be the direction for future work.

\section{Acknowledgements} 

M.E.O.B. was supported by the S{\~a}o Paulo Research Foundation   (FAPESP), grant 2021/03251-0 and V.S.  was supported by the  National Council for Scientific and Technological Development (CNPq) of Brazil,  grant 307813/2019-3.  

\appendix

\section{Suppression functions}

\subsection{Derivation of the generating function}
\label{sup_func}

Now, we will start by explaining in more details the derivation of the generating function used in the main text. We have denoted by $S$ the contingency table with fixed margins for the inputs $\sum^M_{l=1} S_{kl}=m_k$ and outputs $\sum^M_{k=1} S_{kl}=n_l$. Using the multinomial expansion in Eq.(\ref{Gen_fun}), we obtain
\begin{eqnarray}
G_{{\bf m}}( {\bf x}) &=& \sqrt{\mb !} \prod^M_{k=1} \sum_{\sum^M_{l=1}S_{kl}=m_k} \prod^M_{l=1} \frac{(U_{kl} x_l)^{S_{kl}}}{S_{kl}!} \nonumber\\
&=& \sqrt{\mb !} \sum_{S_{kl}\geq 0} \prod^M_{k,l=1} \delta_{\sum^M_{l=1}S_{kl},m_k} \frac{(U_{kl} x_l)^{S_{kl}}}{S_{kl}!} ,
\label{gen_exp}
\end{eqnarray}
Then, replacing Eq.(\ref{gen_exp}) in Eq.(\ref{rec}) we have the following expression
\begin{eqnarray}
_b\langle { \bf n } | { \bf m } \rangle_a &=& 
\sqrt{\frac{\mb !}{\nb !}} \sum_{S_{kl}\geq 0} \left(  \prod^M_{k=1}  \prod^M_{l=1} \delta_{\sum^M_{l=1}S_{kl},m_k} \frac{U_{kl}^{S_{kl}}}{S_{kl}!} \right) \left. \prod^M_{l=1} \frac{\partial^{n_l} }{\partial x^{n_l}_l} x^{\sum^M_{k=1}S_{kl}}_{l}  \right|_{{\bf x} = 0} \nonumber\\
&=& \sqrt{\mb ! \nb !} \sum_{S_{kl}\geq 0}  \prod^M_{k=1}  \prod^M_{l=1} \delta_{\sum^M_{l=1}S_{kl},m_k} \delta_{\sum^M_{k=1}S_{kl},n_l}   \frac{U_{kl}^{S_{kl}}}{S_{kl}!}  ,
\end{eqnarray}
which reduces to the amplitude shown in the main text by denoting $\sum_{ \{ S \}}$ as the sums over $S_{kl}\geq 0$ with the constraints of the margins, and manipulating the factorial elements. In addition, our generating function introduced in Eqs. (\ref{rec}),(\ref{Gen_fun}) depends only in the output formal variables ${\bf x}$ and it is also possible to derive a generating function which depends also in the analogous input formal variables, see Refs. \cite{Jackson, assymp2, Quesada} for instance.

As assumed in the main text, we focus on the mode $l=1$, which can have an arbitrary number of photons $n_1 \geq 1$, and consider that the other modes have few photons. Denoting the output configurations as ${\bf n}=(n_1, {\bf n}_S)$, with  ${\bf n}_S=(n_2,...,n_M)$, we can remove the photons in each mode of ${\bf n}_S$ by using the  recurrence relation  of  Eq. (\ref{recurrence_amp})  repeatedly   $n_l$ times for each output modes $l=2,\ldots,M$. Following this procedure, we obtain  the amplitude $_b\langle {\bf n} | {\bf m} \rangle_a$ as a linear combination of amplitudes in the form
\begin{eqnarray}
_b\langle n_1, {\bm 0}_S | {\bf m}'  \rangle_a &=&
\frac{1}{\sqrt{n_1!}}  \left. \frac{\partial^{n_1}}{\partial x_1^{n_1}}  G_{{\bf m}'} (x_1,0,...,0) \right|_{x_1 = 0} = \sqrt{\frac{n_1!}{{\bf m}'!}} \hspace{0.5mm} U^{m'_k}_{k1} ,
\label{trivial_amp}
\end{eqnarray}
where ${\bf m}' $ is the input configuration with fewer photons that appears in each term of the expansion due to the use of the recurrence relation. Finally, factoring ${\bf m}!$ and the smallest order of $U_{kl}$, i.e. $m_k-|{\bf n}_S|$, we obtain the amplitude in the  form of Eq.(\ref{Form}), where the suppression function $f^{\bf n}_{\bf m}(U)$  is obtained by collecting the matrix elements that appear from the Eqs. (\ref{recurrence_amp}),(\ref{trivial_amp}) and the terms remaining in the factorization. This function is a polynomial in the parameters of the interferometers $\sqrt{\rho}_j$ and $\sqrt{\tau}_j$ and below, will be shown explicitly for the considered cases.

\subsection{Beamsplitter}

First of all, let us consider the simplest suppression laws, which are those with $|{\bf n}_S|=1$. In the main text, it corresponds only to the amplitudes with output configurations ${\bf n}=(n_1, 1)$. In addition, here we also will consider the amplitudes with outputs ${\bf n}=(1, n_2)$. For the first, we need to apply Eq.(\ref{recurrence_amp}) in Eq.(\ref{rec}) for the output mode $l=2$ and for the second, the same with $l=1$,  obtaining Eq.(\ref{Form}) with the respective suppression functions:
\begin{eqnarray}
f^{(n_1,1)} _{(m_1,m_1)} (B) &=&  m_1 B_{12} B_{21} + m_2 B_{11} B_{22} =  (m_1+m_2) \tau - m_1 ,  \\
f^{(1,n_2)} _{(m_1,m_1)} (B) &=& m_1 B_{11} B_{22} + m_2 B_{21} B_{12} =  (m_1+m_2) \tau - m_2 ,
\end{eqnarray}
whose roots coincide with the suppression laws found in Ref. \cite{Jabbour}.  Note that these suppression laws have the same form, differing only by interchanging the input configurations $m_1$ and $m_2$, as expected.

Now, for the amplitudes with with $|{\bf n}_S|=2$, in the main text we have considered the output configuratins with ${\bf n}=(n_1,2)$.  Here, we also will consider the amplitudes with ${\bf n}=(2,n_2)$. Then, for the first we use Eq.(\ref{recurrence_amp}) twice for the mode $l=2$ and for the second, the same for $l=1$, obtaining respectively the following recurrence relations:
\begin{eqnarray}
&& \frac{\partial^2}{\partial^2 x_1} G_{{\bf m}} ({\bf x}) = \nonumber\\
&& = \sqrt{m_1 (m_1-1)} U^2_{11} \hspace{0.5mm} G_{{\bf m} - 2{\bf 1}_1} ({\bf x})+ \sqrt{m_2 (m_2-1)}  U^2_{21} \hspace{0.5mm} G_{{\bf m} - 2{\bf 1}_2 } ({\bf x}) + \nonumber \\ && +  2 \sqrt{m_1 m_2} U_{11} U_{21} \hspace{0.5mm} G_{{\bf m} - {\bf 1}_1 -  {\bf 1}_2 } ({\bf x}) \\
&& \frac{\partial^2}{\partial^2 x_2} G_{{\bf m}} ({\bf x}) = \nonumber\\ &&= \sqrt{m_1(m_1-1)} U^2_{12} \hspace{0.5mm} G_{  {\bf m} - 2{\bf 1}_1 } ({\bf x}) +  \sqrt{m_2(m_2-1)} U^2_{22} \hspace{0.5mm} G_{{\bf m} - 2{\bf 1}_2} ({\bf x}) + \nonumber \\ && +  2 \sqrt{m_1 m_2} U_{12} U_{22} \hspace{0.5mm} G_{{\bf m} - {\bf 1}_1 -  {\bf 1}_2} ({\bf x}) , 
\end{eqnarray}
and replacing in Eq.(\ref{rec}) we obtain Eq.(\ref{Form}) with the suppression functions
 \begin{eqnarray}
&& f^{(n_1,2)} _{(m_1,m_2)} (B) =  \nonumber\\ && = m_1 (m_1-1) B^2_{12} B^2_{21} + 2 m_1 m_2 B_{11} B_{12} B_{21} B_{22}  + m_2 (m_2-1) B^2_{11} B^2_{22}  \nonumber \\ &&= (m_1+m_2)(m_1+m_2-1) \times \nonumber\\ && \times \left[ \tau^2 - \frac{2 m_1}{m_1+m_2} \tau + \frac{m_1(m_1-1)}{(m_1+m_2)(m_1+m_2-1)} \right]  ,
 \label{sup2}
 \\
&& f^{(2,n_2)} _{(m_1,m_2)} (B) =  \nonumber\\ && =  m_1 (m_1-1) B^2_{11} B^2_{22} + 2 m_1 m_2 B_{11} B_{12} B_{21} B_{22}  + m_2 (m_2-1) B^2_{21} B^2_{12}  \nonumber \\ &&= (m_1+m_2)(m_1+m_2-1) \times \nonumber\\ && \times \left[ \tau^2 - \frac{2 m_2}{m_1+m_2} \tau + \frac{m_2(m_2-1)}{(m_1+m_2)(m_1+m_2-1)} \right].
 \label{sup1}
 \end{eqnarray}
Note that, the root of Eq.(\ref{sup2}) is the suppression law shown in Eq.(12) of the main text and the root of Eq.(\ref{sup1}) has the same form of the the previous, but with $m_1$ and $m_2$ interchanged.

\subsection{Tritter}
\label{apptritter}

First of all, for the output configurations with $|{\bf n}_S|=1$, we have ${\bf n}_S=(1,0)$ or ${\bf n}_S=(0,1)$. For the first one, we need to use the recurrence of Eq.(\ref{recurrence_amp}) once for $l=2$, and for the second one, once for $l=3$. Then, replacing at Eq.(\ref{rec}) we obtain (\ref{Form}) with the suppression functions having the following form:
\begin{eqnarray}
&& f^{(n_1,1,0)}_{{\bf m}} (U) = m_1 U_{12} U_{21} U_{31} + m_2 U_{11} U_{22} U_{31} + m_3 U_{11} U_{21} U_{32} ,
\\
&& f^{(n_1,0,1)}_{{\bf m}} (U) = m_1 U_{13} U_{21} U_{31} + m_2 U_{11} U_{23} U_{31} + m_3 U_{11} U_{21} U_{33} .
\end{eqnarray}
Finally, considering our families of tritters $T^{(1)}$ and $T^{(2)}$ as the unitary transformation $U$ of the previous equation and the input configuratins ${\bf m}^{(II)}=(m,m,m)$, we have
\begin{eqnarray}
&& f^{(n_1,1,0)}_{m,m,m} (T^{(1)}) =  - f^{(n_1,0,1)}_{m,m,m} (T^{(1)}) =\frac{m}{3} (2 \tau_1 - 1) ,
\\
&& f^{(n_1,1,0)}_{m,m,m} (T^{(2)}) =   f^{(n_1,0,1)}_{m,m,m} (T^{(2)}) =\frac{m \sqrt{2}}{4} (3 \tau_2 - 2) \sqrt{\tau_2} \text{e}^{i \theta} .
\end{eqnarray}
whose non-trivial roots are $\tau_1=1/2$ or $\tau_2=2/3$, which correspond to the symmetric tritters.  

Now, for $|{\bf n}_S|=2$ we have considered the outputs with ${\bf n}_S=(1,1)$ and ${\bf n}_S=(2,0)$. For the first, we need to use the recurrence of Eq.(\ref{recurrence_amp}) for the modes $l=2$  and $l=3$ simultaneously, for the last we use this recurrence twice for $l=2$, obtaining the following recurrence relations, respectively:
\begin{eqnarray}
&&  \frac{\partial^2}{\partial x_2 \partial x_3} G_{{\bf m}} ({\bf x})  = \nonumber\\
&&  =
\sqrt{m_1 m_2} (U_{12} U_{23} + U_{22} U_{13}) G_{{\bf m} - {\bf 1}_1 -  {\bf 1}_2} ({\bf x}) + \sqrt{m_3(m_3-1)} U_{32} U_{33} G_{{\bf m} - 2{\bf 1}_3} ({\bf x})  + \nonumber\\
&& +
\sqrt{m_1 m_3} (U_{12} U_{33} + U_{32} U_{13}) G_{{\bf m} - {\bf 1}_1 -  {\bf 1}_3} ({\bf x})  +  \sqrt{m_2(m_2-1)} U_{22} U_{23} G_{{\bf m} - 2{\bf 1}_2} ({\bf x})  + \nonumber\\
&& +
\sqrt{m_2 m_3} (U_{22} U_{33} + U_{32} U_{23}) G_{{\bf m} - {\bf 1}_2 -  {\bf 1}_2} ({\bf x})  +  \sqrt{m_1(m_1-1)} U_{12} U_{13} \hspace{0.5mm} G_{{\bf m} - 2{\bf 1}_1} ({\bf x}) , \nonumber\\
\label{sup_trit_n11}
\end{eqnarray}
\begin{eqnarray}
&&  \frac{\partial^2}{\partial^2 x_2} G_{{\bf m}} ({\bf x}) = \nonumber\\
&&  = \sqrt{m_1(m_1-1)} U^2_{12}G_{{\bf m} - 2{\bf 1}_1} ({\bf x})+  \sqrt{m_2(m_2-1)} U^2_{22} \hspace{0.5mm}G_{{\bf m} - 2{\bf 1}_2} ({\bf x}) + \nonumber\\
&& + \sqrt{m_3(m_3-1)} U^2_{32} \hspace{0.5mm} G_{{\bf m} - 2{\bf 1}_3}  ({\bf x})  + 2 \sqrt{m_1 m_2} U_{12} U_{22} \hspace{0.5mm}  G_{{\bf m} - {\bf 1}_1 - {\bf 1}_2}  ({\bf x}) + \nonumber\\
&& + 2 \sqrt{m_1 m_3} U_{12} U_{32} \hspace{0.5mm} G_{{\bf m} - {\bf 1}_1 - {\bf 1}_3}  ({\bf x})  + 2 \sqrt{m_2 m_3} U_{22} U_{32} \hspace{0.5mm} G_{{\bf m} - {\bf 1}_2 - {\bf 1}_3}  ({\bf x})  ,
\end{eqnarray}
and then,  replacing at Eq.(\ref{rec}) we obtain (\ref{Form}) with the corresponding suppression functions:
\begin{eqnarray}
&& f^{(n_1,1,1)} _{{\bf m}} (U) = \nonumber\\
&& =  U_{11} U_{21} U_{31} \Big[ m_1 m_2 \left( U_{12} U_{23} + U_{22} U_{13} \right) U_{31}  +  m_1 m_3 \left( U_{12} U_{33} + U_{32} U_{13} \right) U_{21} +  \nonumber\\ 
&& + m_2 m_3 \left( U_{22} U_{33} + U_{32} U_{23} \right) U_{11} \Big]  + \Big[ m_1 (m_1-1) U_{12} U_{13} U^2_{21} U^2_{31} + \nonumber\\ 
&&  + m_2 (m_2-1) U_{22} U_{23} U^2_{11} U^2_{31} + m_3 (m_3-1) U_{32} U_{33} U^2_{11} U^2_{21} \Big] 
\label{sup_trit_n11}
\end{eqnarray}
\begin{eqnarray}
&& f^{(n_1,2,0)} _{{\bf m}} (U) =  2 U_{11} U_{21} U_{31}   \Big[ m_1 m_2 U_{12} U_{22} U_{31} + m_1 m_3 U_{12} U_{32} U_{21} + \nonumber\\ 
&& +  m_2 m_3 U_{22} U_{32} U_{11} \Big]  + m_1 (m_1 - 1) U^2_{12} U^2_{21} U^2_{31} + \nonumber\\ && + m_2 (m_2 - 1) U^2_{22} U^2_{11} U^2_{31} + m_3 (m_3 -1) U^2_{32} U^2_{11} U^2_{21} ,
\label{sup_trit_n20}
\end{eqnarray}

The previous equation has too many parameters: the input configurations $m_k$, the tritter parameters $\tau_j$ and $\theta$. To find suppression laws it is convenient to consider inputs with only one parameter, in our case ${\bf m}^{(I)}=(n_1,1,1)$ and ${\bf m }^{(II)}=(m,m,m)$, and our families of tritters $T^{(1)}$ and $T^{(2)}$ as the unitary transformation $U$. Then, for which one of these cases, the suppression functions of Eq.(\ref{sup_trit_n11}) are given by:
\begin{eqnarray}
&& f^{(n_1,1,1)}_{(n_1,1,1)} (T^{(1)}) = \nonumber\\ && =  \frac{\sqrt{2}}{18} \Big[ \left( 4  \text{e}^{i 2 \theta} + 3 (1+  \text{e}^{i 2 \theta} ) n_1 + (3- \text{e}^{i 2 \theta} )n^2_1 \right) \tau_1  + 3 n_1 (n_1 - 1 )   \Big] \sqrt{1-\tau_1} ,  \nonumber\\
\end{eqnarray}
\begin{eqnarray}
 && f^{(n_1,1,1)}_{(n_1,1,1)} (T^{(2)}) = \nonumber\\ && = \frac{\sqrt{2}}{4} \Big[ \text{e}^{i 2 \theta} \left( 2 + 3 n_1 + n^2_1 \right) \tau_2 + (3-\text{e}^{i 2 \theta})n_1 - (1+\text{e}^{i 2 \theta}) n_1^2 \Big] \sqrt{\tau_2 (1-\tau_2)} ,  \nonumber\\
\end{eqnarray}
\begin{eqnarray}
&& f^{(n_1,1,1)}_{(m,m,m)} (T^{(1)}) =   \nonumber\\ && = \frac{m}{9} \Big[ 2(2m+ \text{e}^{i 2 \theta}-1)(\tau_1-1)\tau_1 + m -1 \Big] ,
\end{eqnarray}
\begin{eqnarray}
&& f^{(n_1,1,1)}_{(m,m,m)} (T^{(2)}) = \nonumber\\ && + \frac{m}{8} \Big[ 3(3m-1)\text{e}^{i 2 \theta} \tau^2_2-2 \left((6m-2)\text{e}^{i 2 \theta}-1 \right) \tau_2 + (4m-2)\text{e}^{i 2 \theta} - 2  \Big] \tau_2 ,  \nonumber\\
\end{eqnarray}
whose roots give the suppression laws for the amplitudes $_b\langle n_1,1,1 | n_1,1,1  \rangle_a $ and $_b\langle n_1,1,1 | m,m,m  \rangle_a $. These results are shown in blue in Fig. 3 of the main text, where the non-trivial suppression laws are ignored (i.e. those that $\tau_1, \tau_2 = 0,1$). Finally, doing the same the previous one,  we have 
\begin{eqnarray}
&& f^{(n_1,2,0)}_{(n_1,1,1)} (T^{(1)}) = \nonumber\\
&& = \frac{\sqrt{2}}{18} \left[ (4 \text{e}^{i 2 \theta} - 3(\text{e}^{i 2 \theta}-1)n_1-(3+\text{e}^{i 2 \theta})n^2_1) \tau_1 -3n_1(1-n_1) \right] \sqrt{1-\tau_1} + \nonumber \\ 
&& + \frac{\sqrt{6}}{9} i \text{e}^{i \theta} \left[ n_1(2-n_1) - (2+n_1-n^2_1) \tau_1 \right] \sqrt{\tau_1} , 
\label{G1}
\end{eqnarray}
\begin{eqnarray}
 && f^{(n_1,2,0)}_{(n_1,1,1)} (T^{(2)})  = \nonumber\\ && =  \frac{\sqrt{2}}{4} \left[ (2+3n_1+n^2_1) \text{e}^{i 2 \theta} \tau_2 -\left( 3+ \text{e}^{i 2 \theta} +( \text{e}^{i 2 \theta}-1)n_1 \right) n_1 \right] \sqrt{\tau_2 (1-\tau_2)} + \nonumber \\ && + \frac{1}{\sqrt{2}} i \text{e}^{i  \theta} (1-n_1) \left[ n_1-(1+n_1)\tau_2 \right] \sqrt{\tau_2}
 ,
\label{G3}
\end{eqnarray}
\begin{eqnarray}
&& f^{(n_1,2,0)}_{(m,m,m)} (T^{(1)}) =  \nonumber\\ && =  \frac{m}{9} \left[ (4m-2-2 \text{e}^{i 2 \theta})(1-\tau_1)\tau_1-m+1 \right] +  \frac{2m}{27} i  \text{e}^{i \theta} (2 \tau_1-1) \sqrt{3 \tau_1(1-\tau_1)} ,  \nonumber\\
\label{G2}
\end{eqnarray}
\begin{eqnarray}
&& f^{(n_1,2,0)}_{(m,m,m)} (T^{(2)}) =  \nonumber\\ && =  \frac{m}{8} \left[ (9m-3)\text{e}^{i 2 \theta} \tau^2_2  -2\left((6m-2)\text{e}^{i 2 \theta}+1 \right) \tau_2 + 2 \left((2m-1)\text{e}^{i 2 \theta}+1 \right) \right] \tau_2 .  \nonumber\\
\end{eqnarray}
whose roots give the suppression laws for the amplitudes $_b\langle n_1,2,0 | n_1,1,1  \rangle_a $ and $_b\langle n_1,2,0 | m,m,m  \rangle_a $. These results are shown in red in Fig. 3, where the non-trivial suppression laws are also ignored. In addition, some os these suppression laws can be obtained explicitly, which are shown in Table \ref{table1}.

\begin{table} 
\caption{Suppression laws  for tritter}
\begin{tabular}{ccccc}
\br
 & \multicolumn{1}{c}{$\theta=0, \pi$} &  \multicolumn{1}{c}{$\theta=\pm \frac{\pi}{2}$} &   \multicolumn{1}{c}{$\theta=0, \pi$} &  \multicolumn{1}{c}{$\theta=\pm \frac{\pi}{2}$}  \\
 \text{$_b\langle {\bf n} | {\bf m} \rangle_a$} & \multicolumn{1}{c}{$\tau_1$} &  \multicolumn{1}{c}{$\tau_1$} & \multicolumn{1}{c}{$\tau_2$}  & \multicolumn{1}{c}{$\tau_2$}  \\
\mr
 $_b\langle n_1 , 1 , 0 | {\bf m}^{(II)}  \rangle_a$ & $\frac{1}{2}$ & $\frac{1}{2}$ & $\frac{2}{3}$ & $\frac{2}{3}$ \\
$_b\langle n_1 , 1 , 1 | {\bf m}^{(I)}  \rangle_a$ & $\frac{3 n_1 (n_1-1)}{2(n_1+1)(n_1+2)}$ & $\frac{3 n_1}{4(n_1+1)} \hspace{1mm} , \hspace{1mm} n_1 \neq 1$ \footnote{For  $n_1=1$ and $\theta=\pm \pi/2$  there is a suppression  law for the tritter $T^{(1)}$ with  an arbitrary $\tau_1$.}& $\frac{2 n_1 (n_1-1)}{(n_1+1)(n_1+2)}$ & $\frac{4 n_1}{(n_1+1)(n_1+2)}$ \\
$_b\langle n_1 , 2 , 0 | {\bf m}^{(I)} \rangle_a$ & $\frac{1}{2} \hspace{1mm} , \hspace{1mm} n_1=1,2$ & See Eq.(\ref{G1}) & $\frac{2}{3} \hspace{1mm} , \hspace{1mm} n_1=1,2$ &  See Eq.(\ref{G3}) \\
$_b\langle n_1 , 1 , 1 | {\bf m}^{(II)} \rangle_a$ & $\frac{1}{2} \left(1 \pm \frac{1}{\sqrt{m}} \right)$ & $\frac{1}{2} \hspace{1mm}  \hspace{1mm} $ & $\frac{2m-1}{3m-1} \pm \sqrt{\frac{12(4m-1)}{6(3m-1)}}$ & $\frac{2}{3}, \hspace{1mm}  \hspace{1mm} \frac{2m}{3m-1}$  \\
$_b\langle n_1 , 2, 0 | {\bf m}^{(II)} \rangle_a$ & $\frac{1}{2} \hspace{1mm}  \hspace{1mm} $  &  See Eq.(\ref{G2})&  $\frac{2}{3} \hspace{1mm} , \hspace{1mm} \frac{2m}{3m-1}$  & $\frac{2m-1}{3m-1} \pm \sqrt{\frac{12(4m-1)}{6(3m-1)}}$ \\
\br
\end{tabular}
\label{table1}
\end{table}

\section{Suppression laws from permutation symmetry}

In Refs. \cite{Dittel1,Dittel2} were developed the permutation symmetry principle for the derivation of suppression laws. Now we will show that only a part of the suppression laws we found are related to these symmetries. First of all, denoting $S_M$ as the group of permutations of $M$ elements and $\sigma$ their elements, we define the action of the permutation operator $P_\sigma$ in a $M$-dimensional vector as follows
\begin{equation}
P_\sigma 
\begin{pmatrix}
x_1 \\
\vdots \\
x_M
\end{pmatrix} 
=
\begin{pmatrix}
x_{\sigma^{-1}(1)} \\
\vdots \\
x_{\sigma^{-1}(M)} 
\end{pmatrix} .
\end{equation}

Let an input configuration which is symmetric under the operation  $\sigma({\bf m}) = {\bf m}$ and an interferometer $U$ that satisfies:
\begin{equation}
P_\sigma U = Z U \Lambda ,
\label{sym_in}
\end{equation}
where $Z$ is a diagonal unitary matrix related to external phases and $\Lambda$ a diagonal matrix that contains the eigenvectors of $P_\sigma$. Here, we denote $D(d_1, d_1, d_3)$ a being a diagonal matrix with elements $d_1$ ,$d_2$ and $d_3$. Then, according to Refs. \cite{Dittel1,Dittel2}, the outputs ${\bf n}$ satisfying $\lambda^{n_1}_1 ... \lambda^{n_M}_M \neq 1$ are suppressed and considering our choice of input/outputs, the corresponding suppression laws are shown in Table \ref{table_sup} (a), where our tritters are denoted by  $T^{(k)}=T^{(k)}(\tau_k,\theta)$. Similarly, if we have outputs symmetrical under the operation  $\sigma({\bf n}) = {\bf n}$ and an interferometer satisfying
\begin{equation}
U P^{\dagger}_\sigma = \Lambda^* U Z^* ,
\label{sym_out}
\end{equation}
we have suppression for inputs ${\bf m}$ such that $\lambda^{m_1}_1 ... \lambda^{m_M}_M \neq 1$. These suppression laws are shown in Table \ref{table_sup} (b) for our choice of inputs/outputs.

For the interference in a beamsplitter, we need to consider the group $S_2=\{ \mathbb{I}, (12) \}$. From our results, only the suppression laws for the amplitudes  $_b\langle n_1,1 | m,m \rangle_a$ are related to the symmetry principle, since they are zero for $\tau=1/2$, which corresponds to the beamsplitter symmetrical under the permutation $(12)$.

For the interference in a tritter, we need to consider the permutation group $S_3=\{ \mathbb{I}, (12), (13), (23), (123), (132) \}$. From our method, part of the suppression laws obtained for the amplitudes  $_b\langle n_1,2,0 | m,m,m \rangle_a$  are related to the symmetry principle. These amplitudes are zero for the tritter $T_s$, which is symmetric under the permutations $(123)$ and $(321)$, and are related to the constant solutions of Eqs. (\ref{123_1_n20}),(\ref{123_2_n20}). Our tritters also can recover the suppression laws due to the permutations $(12)$ and $(23)$, however, these results are the trivial cases, where some $\tau_j =0$ or $\tau_j =1$. Now, denoting our tritters by $T^{(k)}=T^{(k)}(\tau_k,\theta)$, these last suppression laws are shown in Table \ref{table_sup}.

\begin{table*}[h]
\caption{Suppression laws  for tritter from permutation symmetry}
\begin{tabular}{ccccc}
\br
\multicolumn{5}{c}{a) Output suppression configurations for symmetric inputs $P_\sigma({\bf m})={\bf m}$} \\
 $\sigma$ & $U$  & $\Lambda$ & Eq.(\ref{sym_in}) &  $f^{\bf n} _{\bf m} (U) $  \\
\mr
$(12)$ & $T^{(2)}(1,\theta)$ & $\text{D}(-1,1,1)$ & \makecell{$\langle n_1,1,1| \mb^{(II)} \rangle$ and \\$\langle n_1,2,0|  \mb^{(II)} \rangle$  for odd $n_1$} & \makecell{$\langle 1,1,1| 1,1,1 \rangle$ and \\ $\langle 1,2,0| 1,1,1 \rangle$} \\
$(12)$ & $T^{(2)}(0,0)$ & $\text{D}(1,-1,1)$ &  $\langle n_1,1,1|  \mb^{(II)} \rangle$ for any $n_1$ & Same  \\
$(12)$ & $T^{(2)}(0,\pi)$  & $\text{D}(1,1,-1)$ & $\langle n_1,1,1| \mb^{(II)} \rangle$ for any $n_1$ & Same \\
$(123)$ & $T_s$  & $\text{D}(1,\text{e}^{i2\pi/3},\text{e}^{i4\pi/3})$ & $\langle n_1,2,0|  \mb^{(II)} \rangle$ for any $n_1$ & Same  \\
$(321)$ & $T_s$  & $\text{D}(1,\text{e}^{i4\pi/3},\text{e}^{i2\pi/3})$ &  $\langle n_1,2,0|  \mb^{(II)} \rangle$ for any $n_1$ & Same \\
\br
\multicolumn{5}{c}{b) Input suppression configurations for symmetric outputs $P_\sigma({\bf n})={\bf n}$} \\
 $\sigma$ & $U$  & $\Lambda$ &   Eq.(\ref{sym_out}) & $f^{\bf n} _{\bf m} (U) $ \\
\mr
 $(23)$ & $T^{(1)}(1,\theta)$ &  $\text{D}(1,-1,1)$ & \makecell{$\langle n_1,1,1| \mb^{(I)}  \rangle$ and \\$\langle n_1,1,1|  \mb^{(II)} \rangle$  for any $n_1$} & Same \\
  $(23)$ & $T^{(1)}(0,\theta)$ & $\text{D}(-1,1,1)$ & \makecell{$\langle n_1,1,1| \mb^{(I)}  \rangle$ and \\$\langle n_1,1,1|  \mb^{(II)} \rangle$  for odd $n_1$} &$\langle 1,1,1| 1,1,1  \rangle$  \\
 $(23)$ & $T^{(2)}(1,\theta)$ & $\text{D}(1,1,-1)$ & \makecell{$\langle n_1,1,1|  \mb^{(I)} \rangle$ and \\$\langle n_1,1,1|  \mb^{(II)} \rangle$  for any $n_1$} & Same  \\
\br
\end{tabular}
\label{table_sup}
\end{table*}

\end{document}